# Neutron-proton effective mass splitting and thermal evolution in neutron rich matter

**B. Behera, T. R. Routray** [*]**, S. K. Tripathy**

School of Physics, Sambalpur University, Sambalpur – 768019, Orissa, India

E-mail: trr1@rediffmail.com

**Abstract.** The thermal evolution of properties of neutron rich asymmetric nuclear matter such as entropy density, internal energy density, free energy density and pressure are studied in the non-relativistic mean field theory using finite range effective interactions. In this framework the thermal evolution of nuclear matter properties is directly connected to the neutron and proton effective mass properties. Depending on the magnitude of neutron-proton effective mass splittings, two distinct behaviours in the thermal evolution of nuclear matter properties are noticed.

**1. Introduction**

Temperature dependence of neutron and proton mean fields and equation of state (EOS) of neutron rich asymmetric nuclear matter (ANM) is a subject of current interest for its implications not only in nuclear physics but also in astrophysics. The calculation of EOS of charge neutral beta stable $n + p + e + \mu$ matter at non-zero temperature, $T$, is essential for a better understanding of supernovae matter and the composition and cooling mechanism of newly born neutron stars [1-7]. In recent years, there has been an increasing interest in the study of neutron and proton mean fields and EOS of ANM as well as properties of neutron stars at finite temperature [2, 5-9]. Most of these works have been carried out in the framework of non relativistic mean field theory using effective nucleon-nucleon interactions. In this work we restrict to effective interactions which depend only on the internucleon separation distance '$r$' and the total nucleon density $\rho = \rho_n + \rho_p$ of the medium. Under this formalism the momentum and temperature dependence of neutron and proton mean fields as well as temperature dependence of the interaction part of EOS of ANM are simulated through the finite range exchange parts of the effective interaction $v_{ex}^{l,ul}(r)$ operating between pairs of like ($l$) and unlike ($ul$) nucleons. In view of this, a correct knowledge of the momentum dependence of the neutron and proton mean fields in neutron rich ANM is very crucial while deciding the temperature dependence of various nuclear matter properties. But our limited understanding on this important aspect is revealed from our poor knowledge on the momentum and density dependence of isovector part of nuclear mean field in ANM [10-28].

Theoretical predictions of different models on the momentum dependence of neutron and proton mean fields in ANM can be divided into two distinct groups depending on whether the neutron effective mass ($m_n^*$) goes above that of the proton one ($m_p^*$) [19-24] or the other way around [25-28]. If the finite range exchange interaction $v_{ex}^{l}(r)$ acting between a pair of like nucleons is weak compared to the interaction $v_{ex}^{ul}(r)$ between a pair of unlike nucleons, then $m_n^*$ lies above $m_p^*$. On the other hand if $v_{ex}^{l}(r)$ is strong compared to $v_{ex}^{ul}(r)$, then $m_p^*$ is above $m_n^*$. The controversy on the nature of neutron-proton effective mass splitting, (($m_n^* - m_p^*$) splitting), was in its peak with the prediction of Dirac-Brueckner-Hartree-Fock (DBHF) calculations [25,26] that $m_p^*$ goes above $m_n^*$. However, it was clarified in Refs [22-24] that if the energy dependence of the self-energy is taken into consideration, the DBHF calculation also predicts that $m_n^*$ goes above $m_p^*$. Thus the present status on the nature of ($m_n^* - m_p^*$) splitting is more or less resolved and now it is widely accepted that neutron effective mass goes above that of the proton in neutron rich ANM. However, the magnitude of ($m_n^* - m_p^*$) splitting is yet to be decided [29-34] and it is still an open problem. In view of this, we examine the influence of ($m_n^* - m_p^*$) splitting of different magnitudes on the thermal evolution of various properties of neutron rich ANM in our attempt to look for possible constraints on the uncertainty prevailing on the magnitude of ($m_n^* - m_p^*$) splitting. For this purpose we use a simple density dependent finite range effective interaction. The parameters of the interaction are adjusted [35] to reproduce correct momentum dependence of nuclear mean field in symmetric nuclear matter (SNM) as demanded by optical model fits to nucleon-nucleus scattering data [36-40].

In section-II, we have formulated the EOS of ANM in terms of EOSs of symmetric nuclear matter and pure neutron matter (PNM). In obtaining this we have exploited the isospin symmetry of various quantities in the EOS of ANM alongwith the fact that SNM and PNM constitute the two extremes of ANM. The EOS of ANM thus obtained is worked out for our simple finite range Yukawa interaction. In section-III, thermal evolution of EOS of ANM has been calculated in terms of the thermal evolution of EOSs of SNM and PNM with particular emphasis on the magnitude of ($m_n^* - m_p^*$) splitting in ANM and the results are presented. Section IV contains a brief summary of the work done and conclusions.

## 2. Formalism

The basic quantities in the description of EOS of ANM are the internal energy density $H(\mathbf{r}, Y_p, T)$, entropy density $S(\mathbf{r}, Y_p, T)$, free energy density $F(\mathbf{r}, Y_p, T)$ and pressure $P(\mathbf{r}, Y_p, T)$ which are to be extracted as functions of total nucleon density $\mathbf{r} = \mathbf{r}_n + \mathbf{r}_p$, proton fraction $Y_p = \dfrac{\mathbf{r}_p}{\mathbf{r}}$ and temperature $T$ in any theoretical model. In general, the internal energy density, entropy density, free energy density and pressure have very complicated dependence on $\mathbf{r}, Y_p$ and $T$. However, the isospin symmetry allows to decompose these quantities in a series of even integral powers of $(1 - 2Y_p)$. The results have the crucial advantage that the $Y_p$-dependence of $H(\mathbf{r}, Y_p, T)$, $S(\mathbf{r}, Y_p, T)$, $F(\mathbf{r}, Y_p, T)$ and $P(\mathbf{r}, Y_p, T)$ are separated from their dependence on $r$ and $T$. The contributions coming from higher order terms are found to be small compared to the leading quadratic term in $(1-2Y_p)$ at zero temperature ($T=0$) [41-43] as well as at finite temperature [3,9]. In view of this, for the sake of simplicity we ignore the higher order terms in the infinite series and express these quantities in ANM in the simple forms,

$$H(r,Y_p,T) = H_0(r,T) + (1-2Y_p)^2 H_S(r,T) \quad (1)$$
$$S(r,Y_p,T) = S_0(r,T) + (1-2Y_p)^2 S_S(r,T) \quad (2)$$
$$F(r,Y_p,T) = F_0(r,T) + (1-2Y_p)^2 F_S(r,T) \quad (3)$$
$$P(r,Y_p,T) = P_0(r,T) + (1-2Y_p)^2 F_S(r,T). \quad (4)$$

Further, we note that pure neutron matter and symmetric nuclear matter at same temperature $T$ and total nucleon density $r$ constitute the two extremes of ANM corresponding to $Y_p = 0$ and $Y_p = 1/2$, respectively. In order that equations (1) - (4) be valid in the complete domain of ANM, i.e., from PNM to SNM, it is necessary that $H_0(r,T)$, $S_0(r,Y_p,T)$, $F_0(r,T)$ and $P_0(r,T)$ be identified with the internal energy density, entropy density, free energy density and pressure in SNM. Similarly, $H_s(r,T)$, $S_s(r,T)$, $F_s(r,T)$ and $P_s(r,T)$ are to be identified as the symmetry energy density, symmetry entropy density, free symmetry energy density and symmetry energy pressure, respectively, and defined in terms of the differences between the respective quantities in PNM and SNM,

$$H_S(r,T) = r\, E_S(r,T) = H_n(r,T) - H_0(r,T), \quad (5)$$
$$S_S(r,T) = S_n(r,T) - S_0(r,T) \quad (6)$$
$$F_S(r,T) = F_n(r,T) - F_0(r,T) \quad (7)$$
$$P_S(r,T) = P_n(r,T) - P_0(r,T), \quad (8)$$

where, $E_S(r,T)$ is the nuclear symmetry energy as a functions of $r$ and $T$. The index 'n' in the sub-script refers to the respective quantities in PNM. It is evident from equations (5)-(8) that a complete description of EOS of ANM amounts to separate descriptions of EOSs of PNM and SNM at same temperature $T$ and same total nucleon density $r$. In this work we make use of a simple density dependent finite range effective interaction to calculate the EOSs of SNM and PNM. This interaction was constructed in our earlier work [44] to study momentum and density dependence of neutron and proton mean fields in ANM at zero temperature ($T=0$) and has been used in the finite temperature calculation of EOS of charge neutral $n + p + e + m$ matter in beta equilibrium, i.e., neutron star matter (NSM) [45].

**2.1 *EOSs of SNM and PNM with finite range effective interaction***

The finite range effective interaction used in this work is given by,

$$v_{eff}(\vec{r}) = t_0(1+x_0 P_s)d(\vec{r}) + \frac{1}{6}t_3(1+x_3 P_s)\left[\frac{\vec{r}(R)}{1+b\vec{r}(R)}\right]^g d(\vec{r})$$
$$+ (W + BP_s - HP_t - MP_s P_t) f(r) \quad (9)$$

where, $f(r)$ is the functional form of a short range interaction of conventional form such as Yukawa, Gaussian or exponential and is specified by a single parameter $a$, the range of the interaction. The remaining symbols in equation (9) have their usual meaning. In this work we shall be restricting only to the Yukawa form of $f(r)$.

The energy densities in SNM as well as in PNM at same temperature $T$ and density $\rho$ obtained from the interaction in equation (9) are given by

$$H_0(\rho,T) = \int f_T^{SNM}(\vec{k})(c^2\hbar^2 k^2 + m^2 c^4)^{1/2} d^3k + \frac{(e_0^l + e_0^{ul})}{4} \frac{\rho^2}{\rho_0} + \frac{(e_g^l + e_g^{ul})}{4\rho_0^{g+1}} \left(\frac{\rho}{1+b\rho}\right)^g \rho^2$$

$$+ \frac{(e_{ex}^l + e_{ex}^{ul})}{4\rho_0} \iint f_T^{SNM}(\vec{k}) f_T^{SNM}(\vec{k'}) g_{ex}\left(\left|\vec{k} - \vec{k'}\right|\right) d^3k\, d^3k' \quad (10)$$

and

$$H_n(\rho,T) = \int f_T^{PNM}(\vec{k})(c^2\hbar^2 k^2 + m^2 c^4)^{1/2} d^3k + \frac{e_0^l}{2} \frac{\rho^2}{\rho_0} + \frac{e_g^l}{2\rho_0^{g+1}} \left(\frac{\rho}{1+b\rho}\right)^g \rho^2$$

$$+ \frac{e_{ex}^l}{2\rho_0} \iint f_T^{PNM}(\vec{k}) f_T^{PNM}(\vec{k'}) g_{ex}\left(\left|\vec{k} - \vec{k'}\right|\right) d^3k\, d^3k'. \quad (11)$$

The new parameters $e_0^i$, $e_g^i$ and $e_{ex}^i$ with $i = l, ul$ appearing in these equations can be connected with the parameters of the interaction [46]. The functional $g_{ex}\left(\left|\vec{k} - \vec{k'}\right|\right)$ is the normalized Fourier transformation of the Yukawa interaction. The Fermi-Dirac momentum distribution functions $f_T^{SNM}(\vec{k})$ and $f_T^{PNM}(\vec{k})$, are defined as

$$f_T^i(\vec{k}) = \frac{x}{(2\pi)^3} n_T^i(k), \quad (12.a)$$

where, $n_T^i(k)$, with $i = 0, n$ are the occupation probabilities,

$$n_T^i(\vec{k}) = \frac{1}{\exp\{[\in^i(k,\rho,T) - \mu^i(\rho,T)]/T\} + 1} \quad (12.b)$$

in SNM and PNM, respectively. The distribution functions are subject to the constraint $\int f_T^{SNM(PNM)}(\vec{k}) d^3k = \rho$. Here, the spin-isospin degeneracy factor $x$ takes value 4(2) in SNM (PNM). In equation (12.b) $\in^i$ and $\mu^i$, with $i = 0, n$ are the respective single particle energies and chemical potentials.

The single particle energies in SNM and PNM can be obtained as the functional derivatives of respective energy densities in equations (10) and (11) and are given by,

$$\in^0(k,\rho,T) = (c^2\hbar^2 k^2 + m^2 c^4)^{1/2} + \frac{(e_0^l + e_0^{ul})}{2} \frac{\rho}{\rho_0} + \frac{(e_g^l + e_g^{ul})}{2\rho_0^{g+1}} \left(\frac{\rho}{1+b\rho}\right)^{g+1} \left(1 + b\rho + \frac{g}{2}\right)$$

$$+ \frac{(e_{ex}^l + e_{ex}^{ul})}{2\rho_0} \int f_T^{SNM}(\vec{k'}) g_{ex}\left(\left|\vec{k} - \vec{k'}\right|\right) d^3k' \quad (13)$$

and

$$\in^n(k,\rho,T) = (c^2\hbar^2 k^2 + m^2 c^4)^{1/2} + e_0^l \frac{\rho}{\rho_0} + \frac{e_g^l}{\rho_0^{g+1}} \left(\frac{\rho}{1+b\rho}\right)^{g+1} \left(1 + b\rho + \frac{g}{2}\right)$$

$$+ \frac{e_{ex}^l}{\rho_0} \int f_T^{PNM}(\vec{k'}) g_{ex}\left(\left|\vec{k} - \vec{k'}\right|\right) d^3k'. \quad (14)$$

It is evident from the expressions of energy density and single particle energy in both SNM and PNM in equations (10), (11), (13) and (14) that the temperature dependence of the mean fields and the interaction parts of energy densities are simulated through the respective Fermi-Dirac momentum distribution functions while the interaction itself has no explicit dependence on temperature. The momentum dependent parts of the mean fields involve the respective distribution functions and therefore imply self-consistent calculations. The momentum distribution functions as well as mean fields both in SNM and PNM can be evaluated self-consistently at each density $r$ and temperature $T$ by adopting an iterative procedure [47]. Here the basic input is the respective single particle energies at zero temperature. At $T=0$, the momentum distribution functions take the form of step functions and the respective mean fields as well as the complete EOSs of SNM and PNM can be calculated analytically. The analytical results of single particle energies in SNM and PNM at zero-temperature, $T=0$, for the interaction in equation (9) are,

$$\in^0 (k, r, T=0) = \left(c^2 \hbar^2 k^2 + m^2 c^4\right)^{1/2} + \frac{(e_0^l + e_0^{ul})}{2}\frac{r}{r_0} + \frac{(e_g^l + e_g^{ul})}{2 r_0^{g+1}} \left(\frac{r}{1+br}\right)^{g+1} \left(1+bg+\frac{g}{2}\right)$$

$$+ \frac{(e_{ex}^l + e_{ex}^{ul})r}{2r_0} \left[ \begin{array}{c} \frac{3\Lambda^2 \left(\Lambda^2 + k_f^2 - k^2\right)}{8kk_f^3} \ln\left\{\frac{\Lambda^2 + (k+k_f)^2}{\Lambda^2 + (k-k_f)^2}\right\} + \frac{3\Lambda^2}{2k_f^2} \\ -\frac{3\Lambda^3}{2k_f^3}\left\{\tan^{-1}\left(\frac{k+k_f}{\Lambda}\right) - \tan^{-1}\left(\frac{k-k_f}{\Lambda}\right)\right\} \end{array} \right]$$

(15)

and

$$\in^n (k, r, T=0) = \left(c^2 \hbar^2 k^2 + m^2 c^4\right)^{1/2} + e_0^l \frac{r}{r_0} + \frac{e_g^l}{r_0^{g+1}} \left(\frac{r}{1+br}\right)^{g+1} \left(1+bg+\frac{g}{2}\right)$$

$$+ e_{ex}^l \frac{r}{r_0} \left[ \begin{array}{c} \frac{3\Lambda^2 \left(\Lambda^2 + k_n^2 - k^2\right)}{8kk_n^3} \ln\left\{\frac{\Lambda^2 + (k+k_n)^2}{\Lambda^2 + (k-k_n)^2}\right\} + \frac{3\Lambda^2}{2k_n^2} \\ -\frac{3\Lambda^3}{2k_n^3}\left\{\tan^{-1}\left(\frac{k+k_n}{\Lambda}\right) - \tan^{-1}\left(\frac{k-k_n}{\Lambda}\right)\right\} \end{array} \right]$$

(16)

where, $\Lambda = \frac{1}{a}$. The Fermi momenta $k_f$ and $k_n$ are related to the density $r$ as $k_f^3 = \frac{3p^2 r}{2}$ and $k_n^3 = 3p^2 r$. The temperature dependence of single particle energies and chemical potentials in SNM and PNM at given density $r$ and temperature $T$ are obtained in the process of self-consistent evaluation of the respective momentum distribution functions. Once the momentum distribution functions, single particle energies and chemical potentials are obtained, the entropy density, internal energy density, free energy density, pressure, etc., in SNM and in PNM can be calculated by adopting standard procedure.

**3. Thermal Evolution of Nuclear Matter Properties**

From the expressions of energy densities and mean fields both in SNM and PNM given in equations (10), (11), (13) and (14) in the previous section, it is evident that the temperature and momentum dependence of the respective mean fields and temperature dependence of the interaction parts of corresponding EOSs are simulated by finite range exchange interactions operating between pairs of like and unlike nucleons. In view of this, thermal evolution of EOSs of SNM and PNM, i.e., properties of SNM and PNM relative to their zero-temperature results, can be calculated only in terms of the exchange parameters $e_{ex}^{l}$, $e_{ex}^{ul}$ and the range $a$ without having to require the knowledge of other parameters $e_{0}^{l}$, $e_{0}^{ul}$, $e_{g}^{l}$, $e_{g}^{ul}$, $b$ and the exponent $g$. In view of this the momentum distribution functions both in SNM and PNM can be obtained self-consistently at each nucleon density $\rho$ and temperature $T$ by introducing effective single particle energies in SNM and PNM which contain only the momentum dependent parts and can be written as,

$$\epsilon_{eff}^{0}(k,\rho,T) = \left[ (c^2\hbar^2 k^2 + m^2 c^4)^{1/2} + \frac{(e_{ex}^{l} + e_{ex}^{ul})}{2\rho_0} \int f_T^{SNM}(\vec{k'}) g_{ex}\left(\left|\vec{k}-\vec{k'}\right|\right) d^3 k' \right], \quad (17)$$

and

$$\epsilon_{eff}^{n}(k,\rho,T) = \left[ (c^2\hbar^2 k^2 + m^2 c^4)^{1/2} + \frac{e_{ex}^{l}}{\rho_0} \int f_T^{PNM}(\vec{k'}) g_{ex}\left(\left|\vec{k}-\vec{k'}\right|\right) d^3 k' \right]. \quad (18)$$

These effective single particle energies would correspond to the respective effective chemical potentials $\mu_{eff}^{0}(\rho,T)$ and $\mu_{eff}^{n}(\rho,T)$ so that $\left[\epsilon^{i}(k,\rho,T) - \mu^{i}(\rho,T)\right] = \left[\epsilon_{eff}^{i}(k,\rho,T) - \mu_{eff}^{i}(\rho,T)\right]$ with $i = 0, n$.

The study of thermal evolution of different properties of SNM and PNM in case of momentum dependent mean fields obtained for the interaction in equation (9) now requires only the knowledge of the parameters $e_{ex}^{l}$, $e_{ex}^{ul}$ and $a$. The parameters $a$ and $e_{ex} = (e_{ex}^{l} + e_{ex}^{ul})/2$ appearing in the effective single particle energy in SNM must be determined so as to provide a correct momentum dependence of the mean field in SNM at normal density $\rho_0$ and at zero-temperature as demanded by optical model fits to nucleon-nucleus scattering data at intermediate energies [36-40]. The values of these two parameters obtained through an optimization procedure for the Yukawa form of finite range exchange interactions are $e_{ex} = (e_{ex}^{l} + e_{ex}^{ul})/2 = -121.8$ MeV and $a = 0.4044$ fm [47]. In obtaining these parameters we have used only the standard values of $mc^2 = 939$ MeV, energy per nucleon at normal density $\frac{H_0(\rho = \rho_0, T = 0)}{\rho_0} = 923$ MeV and $(c^2\hbar^2 k_{f_0}^2 + m^2 c^4)^{1/2} = 976$ MeV (corresponding to $\rho_0 = 0.1658$ fm$^{-3}$). The effective nucleon mass $m^*(k = k_{f_0}, \rho_0, T = 0)/m$ in SNM at normal density and $T = 0$ is found to be 0.67. The momentum dependence of the mean field in SNM at zero-temperature calculated with these parameters is in good agreement over a wide range of density and momentum with the results obtained by Wiringa [48] using realistic interaction UV14+UVII as has been shown in Ref.[47]. Once the parameters $(e_{ex}^{l} + e_{ex}^{ul})/2$ and $a$ are fixed, the thermal evolution of properties of SNM can be calculated. However, to obtain the thermal evolution of properties of PNM we require the splitting of the combination $(e_{ex}^{l} + e_{ex}^{ul})$ into $e_{ex}^{l}$ and $e_{ex}^{ul}$ for interactions between pairs of like and unlike nucleons. In this context we note that if the parameter $e_{ex}^{l}$ is chosen in between 0 and $e_{ex}$, then $m_n^*$ in neutron rich ANM will go above $m_p^*$. On the other hand, $m_p^*$ will go above $m_n^*$ if $e_{ex}^{l}$ lies between $e_{ex}$ and $2 e_{ex}$ [44]. This controversy in connection with $(m_n^* - m_p^*)$ splitting in neutron rich ANM is more or less resolved and now it is widely accepted that neutron effective mass goes above the proton one. However, there is no consensus among various theoretical models on the magnitude of $(m_n^* - m_p^*)$ splitting. In view of this while the parameter $e_{ex}^l$ should lie in between 0 and $e_{ex}$, its actual value is quite uncertain. Different choices of $e_{ex}^l$ in the range 0 and $e_{ex}$ will therefore predict different magnitudes of $(m_n^* - m_p^*)$ splittings leading to different thermal evolution of various properties in PNM. Under this circumstance, we vary the parameter $e_{ex}^l$ in the range 0 to $e_{ex}$ and examine its influence on the thermal evolution of different properties in PNM. The results in SNM and PNM thus obtained are also compared with the respective predictions of ideal Fermi gas model where the mean fields are momentum independent. The momentum distribution functions in the ideal Fermi gas model are evaluated by considering only the kinetic energy terms in the expressions of the effective single particle energies in equations (17) and (18), and the thermal evolution of SNM and PNM are calculated. Comparison of the results of our interaction in Eq.(9) with the predictions of ideal Fermi gas model will bring out the effect of momentum dependence of the nuclear mean field on the thermal evolution of various properties. In this context we note that the specific choice $e_{ex}^l = 0$ corresponds to a momentum independent mean field in PNM and as a result thermal evolution of various properties in this case will be identical to those obtained in the ideal Fermi gas model.

Once the momentum distribution functions $f_T^i(\vec{k}), i=0,n$, are obtained self-consistently in SNM and PNM, the respective entropy densities, $S_i(r,T)$ can be calculated from the relation,

$$S_{0,n}(r,T) = -\frac{x}{(2\pi)^3} \int \left[ n_T^{0,n}(k) \ln n_T^{0,n}(k) + \left(1 - n_T^{0,n}(k)\right) \ln\left(1 - n_T^{0,n}(k)\right) \right] d^3k, \quad (19)$$

where, $x = 4(2)$ for SNM (PNM) and $n_T^{0,n}(k)$ are the respective occupation probabilities at temperature $T$. The results of entropy densities in SNM and the four different cases of PNM, namely, $e_{ex}^l = 0, e_{ex}/3, 2e_{ex}/3$ and $e_{ex}$, are shown as a function of density $r$ at two different temperatures, $T = 40$ and 60 MeV, in figures 1(a) and (b) respectively. The entropy density $S_0(r,T)$ in SNM calculated in the ideal Fermi gas model are also shown in these two figures. Comparing our results with those obtained from ideal Fermi gas model in both SNM and PNM it is found that the role of momentum dependence of the mean fields is to reduce the effect of temperature. It is also noticed from the figures that for a given temperature $T$, the entropy density obtained in the ideal Fermi gas model in SNM goes above that of PNM and the difference increases with density. On the other hand our results of entropy density in PNM exceeds that of SNM at a higher density if the parameter $e_{ex}^l$ lies between 0 and $2e_{ex}/3$, which is contrary to the result obtained with ideal Fermi gas model. The density at which the entropy density in PNM exceeds that of SNM gradually increases with increase in the magnitude of the parameter $e_{ex}^l$ from 0 to $2e_{ex}/3$. For the specific choice of $e_{ex}^l = 2e_{ex}/3$, the entropy density in PNM approaches that of SNM asymptotically at large density. With increase in the magnitude of $e_{ex}^l$ from the critical value $2e_{ex}/3$ towards $e_{ex}$, the curves of entropy density in PNM are pushed more and more below that of SNM, a behaviour similar to results obtained in the ideal Fermi gas model. The question which, now, automatically arises is, whether the entropy density in PNM being a system of one kind of particle can exceed that of SNM which is a two component system or not. A plausible answer to this pertinent question may be searched in the areas of study of heavy-ion collision dynamics involving highly neutron rich radioactive nuclei. If an answer to this basic question can be obtained then accordingly the magnitude of $(m_n^* - m_p^*)$ splitting in ANM can be further constrained.

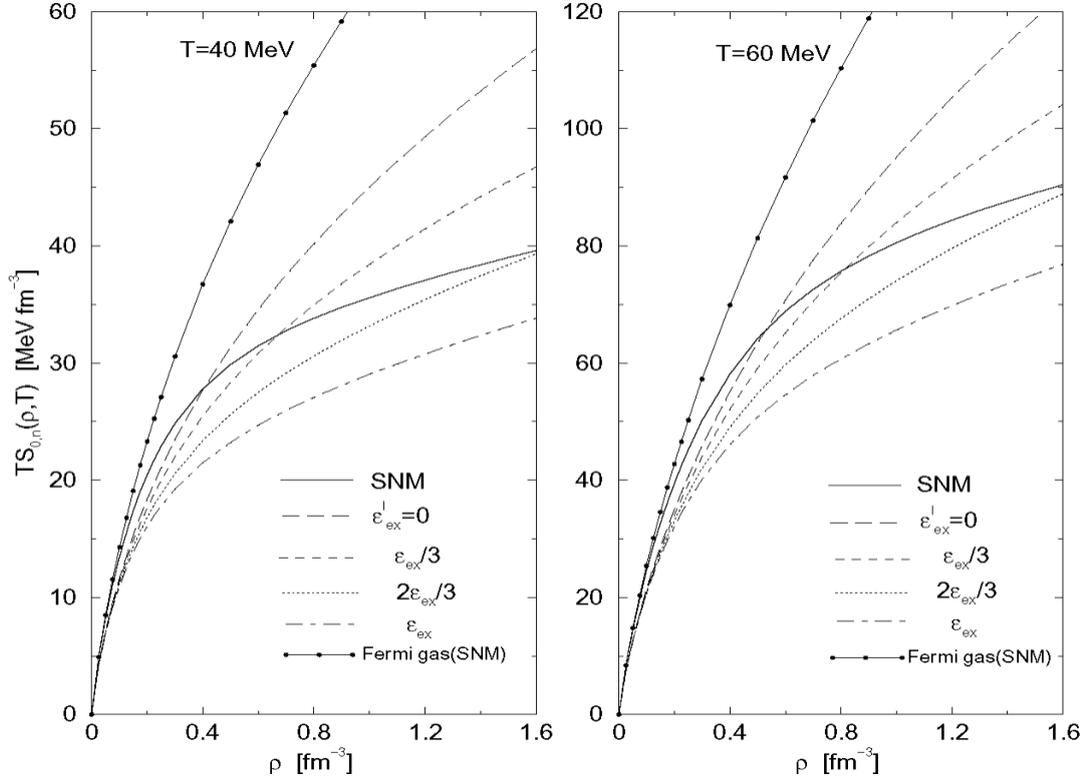

**Figures 1(a) & (b).** Entropy density, $TS_{0,n}(r,T)$, in SNM and four cases of PNM corresponding to $\varepsilon_{ex}^l = 0$, $\varepsilon_{ex}/3$, $2\varepsilon_{ex}/3$ and $\varepsilon_{ex}$ shown as a function of density $r$ at temperature T= 40 MeV. Fermi-gas model results in SNM and PNM correspond to $\varepsilon_{ex} = 0$ and $\varepsilon_{ex}^l = 0$, respectively. (b) Same as (a) but at temperature, T=60 MeV.

We shall now examine the thermal evolution of other properties of SNM and PNM, such as, internal energy density, free energy density and pressure for the different choices of $\varepsilon_{ex}^l$ within its allowed range. The thermal evolution of internal energy density, $H_{0(n)}^{th}(r,T)$, in SNM (PNM) can be given by,

$$H_{0(n)}^{th}(r,T) = \frac{x}{(2\pi)^3}\left[\int \left(c^2\hbar^2 k^2 + M^2 c^4\right)^{1/2} n_T^{0(n)}(\vec{k})\, d^3k - \int_{k_{f(n)}} \left(c^2\hbar^2 k^2 + M^2 c^4\right)^{1/2} d^3k\right],$$
$$+ \frac{A}{2}\left(\frac{x}{(2\pi)^3}\right)^2 \left[\iint n_T^{0(n)}(\vec{k})\, n_T^{0(n)}(\vec{k}')\, g_{ex}(|\vec{k}-\vec{k}'|)\, d^3k\, d^3k' - \iint_{k_{f(n)}} g_{ex}(|\vec{k}-\vec{k}'|)\, d^3k\, d^3k'\right]$$
(20)

where, $A = \frac{(e_{ex}^{l} + e_{ex}^{ul})}{2r_0}\left(\frac{e_{ex}^{l}}{r_0}\right)$ and $x = 4(2)$. The integral $\int_{k_{f(n)}}$ implies integration over the Fermi sphere of radius $k_{f(n)}$ in SNM (PNM). As mentioned earlier, equation (20) shows that only the kinetic energy terms and finite range exchange terms contribute to the thermal evolution of nuclear properties in SNM (PNM). The results of thermal evolution of internal energy density in SNM and the four different cases of PNM are shown in figure 2 as functions of density $r$, at temperature $T = 40$ MeV. The ideal Fermi gas model result in SNM is also given in the same figure for comparison. The thermal evolution of free energy density in SNM (PNM) can now be given in terms of entropy density, $TS_{0(n)}(r,T)$ and $H_{0(n)}^{th}(r,T)$ as $F_{0(n)}^{th}(r,T) = H_{0(n)}^{th}(r,T) - TS_{0(n)}(r,T)$. The density dependence of $F_{0(n)}^{th}(r,T)$ in SNM (PNM) is calculated at temperature $T = 40$ MeV for the same cases as in figure 2, and the results are shown in figure 3. The thermal evolution of pressure can be given by, $P_{0(n)}^{th}(r,T) = r[m^{0(n)}(r,T) - m^{0(n)}(r,T=0)] - F_{0(n)}^{th}(r,T)$. From the definition of effective chemical potentials, it is apparent that $[m_{eff}^{0(n)}(r,T) - m_{eff}^{0(n)}(r,T=0)] = [m^{0(n)}(r,T) - m^{0(n)}(r,T=0)]$. The effective chemical potential in case of SNM (PNM) at finite temperature is obtained in the process of self-consistent evaluation of momentum distribution function $f_T^{SNM(PNM)}(\vec{k})$, whereas, its zero-temperature counterpart is obtained from equation (15) (equation (16)) by considering only the kinetic and finite range exchange terms evaluated at the Fermi momentum $k = k_{f(n)}$ corresponding to density $r$. The thermal evolution of pressure, $P_{0(n)}^{th}(r,T)$, in SNM (PNM) as function of density at $T = 40$ MeV is given in figure 4 for the same cases, as in figure 3.

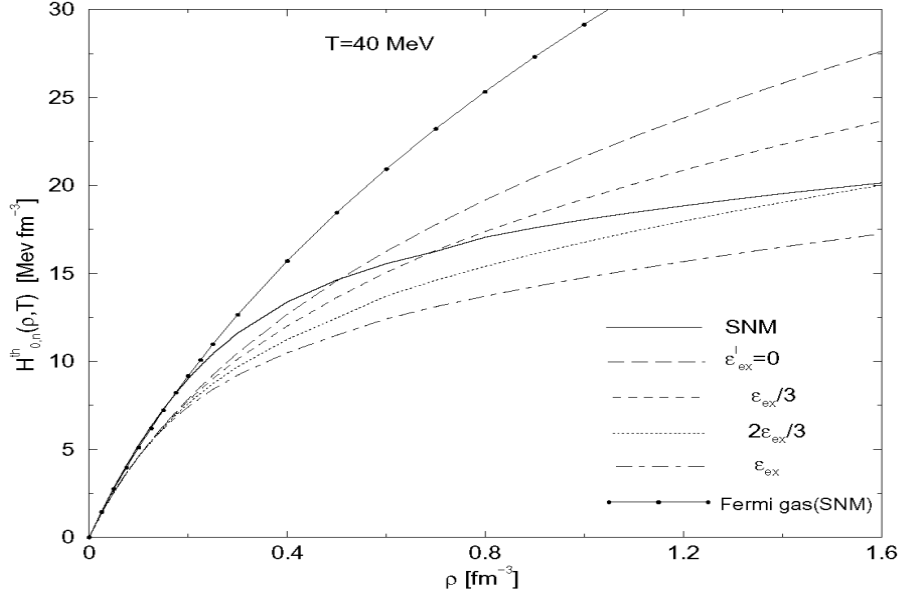

**Figure 2** Thermal evolution of internal energy density, $H^{th}_{0,n}(r,T)$, in SNM and four cases of PNM corresponding to $e^l_{ex}=0, e_{ex}/3, 2e_{ex}/3$ and $e_{ex}$ shown as a function of density $r$ at temperature T= 40 MeV. Fermi-gas model results in SNM and PNM correspond to $e_{ex}=0$ and $e^l_{ex}=0$, respectively.

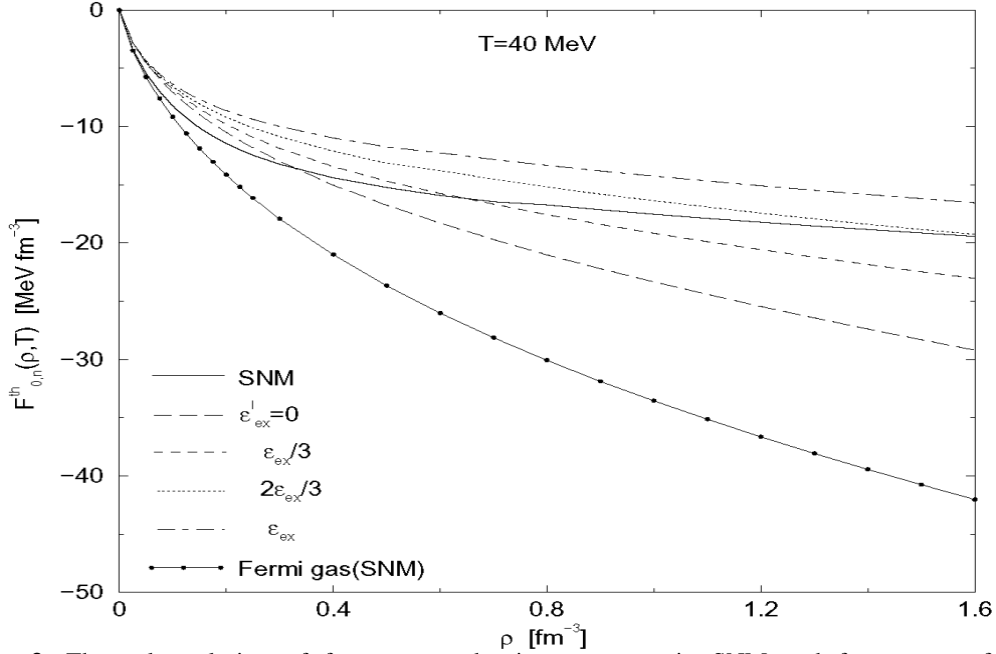

**Figure 3.** Thermal evolution of free energy density, $F^{th}_{0,n}(r,T)$, in SNM and four cases of PNM corresponding to $e^l_{ex}=0, e_{ex}/3, 2e_{ex}/3$ and $e_{ex}$ shown as a function of density $r$ at temperature T= 40 MeV. Fermi-gas model results in SNM and PNM correspond to $e_{ex}=0$ and $e^l_{ex}=0$, respectively.

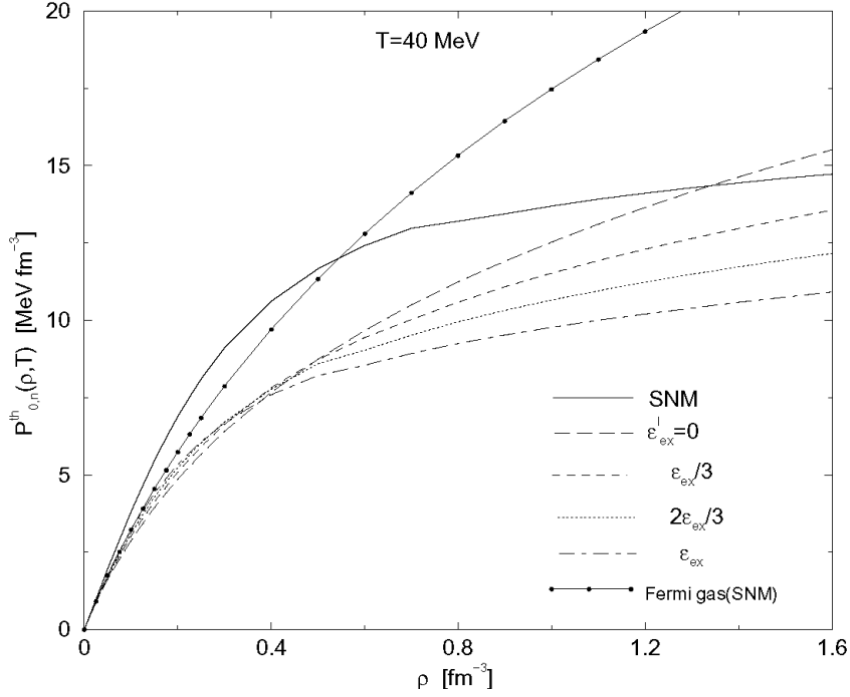

**Figure 4.** Thermal evolution of pressure, $P_{0,n}^{th}(\rho,T)$, in SNM and four cases of PNM corresponding to $\varepsilon_{ex}^{l}=0$, $\varepsilon_{ex}/3$, $2\varepsilon_{ex}/3$ and $\varepsilon_{ex}$ shown as a function of density $\rho$ at temperature T= 40 MeV. Fermi-gas model results in SNM and PNM correspond to $\varepsilon_{ex}=0$ and $\varepsilon_{ex}^{l}=0$, respectively.

    The thermal evolutions of internal energy density and pressure have positive values both in SNM and PNM, whereas, it is negative in case of free energy density as can be seen from respective figures (2), (4) and (3). It can also be seen from figures (2) and (3) that thermal evolutions of internal energy density and free energy density for different choices of $\varepsilon_{ex}^{l}$ in PNM show similar behaviour relative to the respective results in SNM as in case of entropy density in figure 1. For $\varepsilon_{ex}^{l}$ in the range 0 to $2\varepsilon_{ex}/3$ the curves of PNM crossover that of SNM at certain higher densities, whereas, there is no such crossing over of PNM and SNM curves at any density in case $\varepsilon_{ex}^{l}$ is in between $2\varepsilon_{ex}/3$ and $\varepsilon_{ex}$. For the critical case of $\varepsilon_{ex}^{l}=2\varepsilon_{ex}/3$, the thermal evolution of internal energy density and free energy density in PNM approach the SNM result asymptotically. However, in case of thermal evolution of pressure, deviation from this common behaviour occurs as can be seen from figure 4. The deviation from this common behaviour is due to the thermal evolution of chemical potential term, $[\mu^{0(n)}(\rho,T)-\mu^{0(n)}(\rho,T=0)]$. We shall now examine the thermal evolution of nuclear symmetry energy and free symmetry energy which are two important quantities in the studies of formation and cooling mechanism of neutron stars.

### 3.1 *Thermal Evolution of Symmetry energy and Free symmetry Energy*

    The thermal evolution of symmetry energy $Q(\rho,T)$ and free symmetry energy $Q_f(\rho,T)$ can be obtained in terms of thermal evolutions of internal energy densities and free energy densities in PNM and SNM and can be expressed as,

$$Q(r,T) = \frac{H_n^{th}(r,T) - H_0^{th}(r,T=0)}{r} \tag{21}$$

and
$$Q_f(r,T) = \frac{F_n^{th}(r,T) - F_0^{th}(r,T=0)}{r} \tag{22}$$

The thermal evolutions of these quantities are calculated as a function of density for different values of $e_{ex}^l$ at temperature $T=40$ MeV and the results are shown in figures 5 and 6 respectively. The results of $Q(r,T)$ and $Q_f(r,T)$ obtained in the ideal Fermi gas model are also shown in these figures for comparison. Thermal evolution of symmetry energy $Q(r,T)$ is negative at low density and decreases with increase in density, attains a minimum and then increases. On the other hand thermal evolution of free symmetry energy $Q_f(r,T)$ has a large positive value at very low density and it decreases with increase in density. A changeover in sign at certain higher density occurs in both $Q(r,T)$ and $Q_f(r,T)$ if $e_{ex}^l$ is chosen in the range 0 to $2e_{ex}/3$. This is also the expected behaviour as can be seen from the thermal evolutions of internal energy density and free energy density in SNM and PNM in figures 2 and 3 respectively. For the critical case of $e_{ex}^l = 2e_{ex}/3$ the thermal evolutions of these quantities in PNM at best approach that of SNM thereby implying vanishing of $Q(r,T)$ and $Q_f(r,T)$ in the asymptotic region of large density which can be seen from figures 5 and 6. On the other hand, if the value of $e_{ex}^l$ lies in between $2e_{ex}/3$ and $e_{ex}$ then no change of sign take place in the functionals $Q(r,T)$ and $Q_f(r,T)$ which is similar to the results obtained in the ideal Fermi gas model.

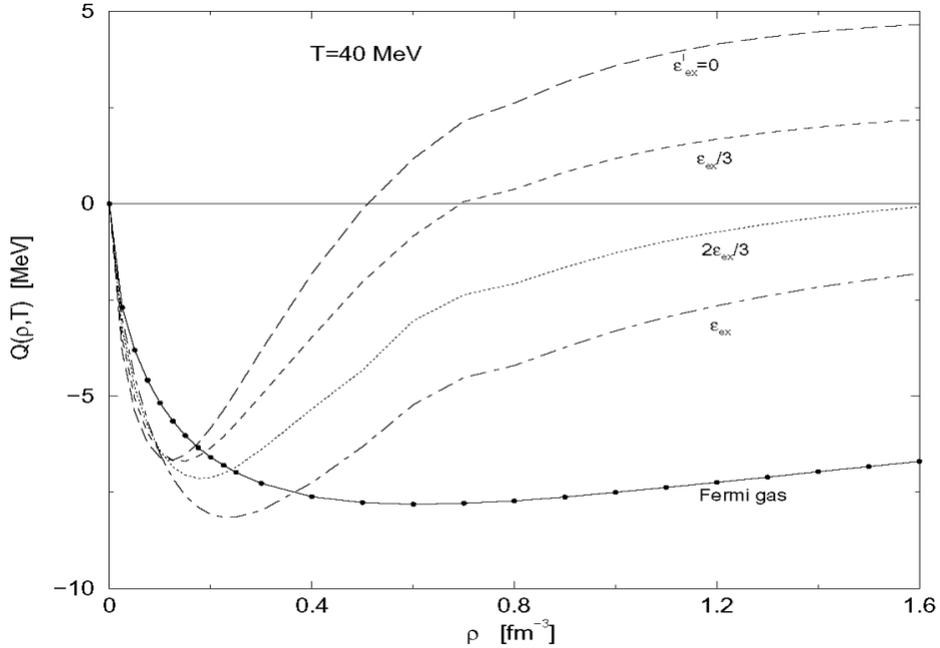

**Figure 5.** Thermal evolution of symmetry energy, $Q(r,T)$, shown as a function of density $r$ at temperature T= 40 MeV for four cases of $e_{ex}^l = 0$, $e_{ex}/3$, $2e_{ex}/3$ and $e_{ex}$. The Fermi gas model result is given as solid line with filled circles.

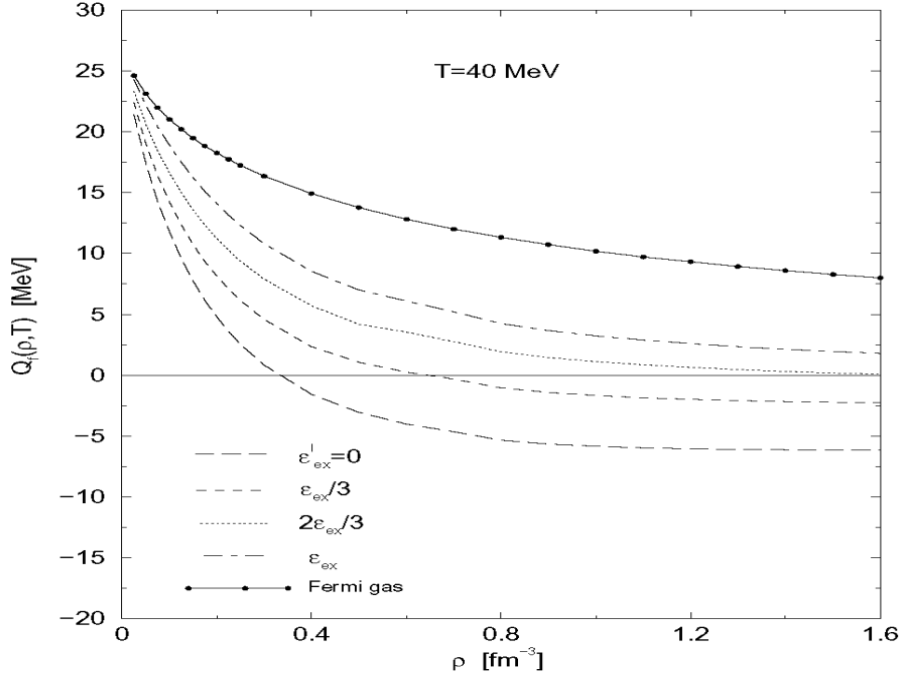

**Figure 6.** Thermal evolution of free symmetry energy, $Q_f(r,T)$, shown as a function of density $r$ at temperature T= 40 MeV for four cases of $e_{ex}^l = 0$, $e_{ex}/3$, $2e_{ex}/3$ and $e_{ex}$. The Fermi gas model result is given as solid line with filled circles.

Two distinct behaviours in the thermal evolution of nuclear matter properties in PNM relative to their respective SNM results are found corresponding to the value of $e_{ex}^l$ in the ranges, (a) 0 to $2e_{ex}/3$ and (b) $2e_{ex}/3$ to $e_{ex}$. The results of thermal evolution of these properties in ANM will be in between the corresponding results in SNM and PNM, as can be seen from equations (1)-(4), and shall correspond to the same characteristic behaviour depending on the choice of $e_{ex}^l$ in these two ranges. We shall now examine the neutron and proton effective mass splitting in ANM at zero-temperature for different choices of $e_{ex}^l$ in the whole range between 0 and $e_{ex}$. The calculation of $m_n^*$ and $m_p^*$ in ANM requires only the momentum dependent parts of the mean fields which are obtained from the finite range exchange interactions. For different combinations of $e_{ex}^l$ and $e_{ex}^{ul}$, the effective masses $m_n^*$ and $m_p^*$ in ANM are calculated as a function of asymmetry $b = (1-2Y_P)$ at normal density $r_0$. It is found that the magnitude of $(m_n^* - m_p^*)$ splitting is maximum at $e_{ex}^l = 0$ and it decreases as the value of $e_{ex}^l$ approaches towards $e_{ex}$ and ultimately vanishes at $e_{ex}^l = e_{ex}$. In figure 7, the results of $(m_n^* - m_p^*)$ splitting obtained for the two specific choices, namely, $e_{ex}^l = e_{ex}/3$ and the critical case $e_{ex}^l = 2e_{ex}/3$ are given. The magnitude of $(m_n^* - m_p^*)$ splitting obtained for the two different choices of $e_{ex}^l$ are compared with the microscopic results of the DBHF calculation using Bonn B potential [49] and results of Bueckner-Hartree-Fock (BHF) approximation to Bueckner-Bethe-Goldstone (BBG) theory [19] obtained using the realistic Argonne $v_{18}$ interaction. The BHF calculation with inclusion of rearrangement term and renormalization contributions is referred to as extended BHF (EBHF) [20]. In the EBHF and BHF calculations the phenomenological three body contributions (3-BF) have been included in order to obtain saturation [21]. The results of $(m_n^* - m_p^*)$ splitting for the critical case $e_{ex}^l = 2e_{ex}/3$ compares well with the microscopic DBHF prediction, whereas, BHF+3-BF and EBHF+3-BF calculations predict larger splittings which will correspond to a smaller value of $e_{ex}^l$ in comparison to $e_{ex}^l = 2e_{ex}/3$. The difference between the curves of $m_n^*/m$ (as well as those of $m_p^*/m$) appearing in the figure for different theoretical models are mainly due to the differences in the effective masses in SNM.

The $(m_n^* - m_p^*)$ splitting in ANM is determined by the momentum dependence of isovector part of the mean field, $u_t(k,\mathbf{r})$. The momentum dependence of $u_t(k,\mathbf{r})$ can be studied with the knowledge of finite range exchange part of the interaction (i.e, by knowing the parameters $e_{ex}^l$, $e_{ex}^{ul}$ and $a$), density dependence of symmetry energy, $E_s(\mathbf{r})$, and effective mass in SNM from the relation [44],

$$u_t(k,\mathbf{r}) = E_s(\mathbf{r}) - \frac{\hbar^2 k_f^2}{3m}\left[\left(\frac{m^*(k,\mathbf{r})}{m}\right)^2 + \frac{\hbar^2 k^2}{m^2 C^2}\right]_{k=k_f}^{-1/2} + u_t^{ex}(k,\mathbf{r}), \qquad (23)$$

with, $\qquad u_t^{ex}(k,\mathbf{r}) = \frac{\mathbf{r}}{2}(e_{ex}^l - e_{ex}^{ul})\int [j_0(kr) - j_0(k_f r)] j_0(k_f r) f(r) d^3r, \qquad (24)$

where, $j_0$ is zeroth order Bessel function. Momentum dependence of $u_t(k,\mathbf{r}_0)$ at normal density $\mathbf{r}_0$ has been studied for the two cases, $e_{ex}^l = e_{ex}/3$ and the critical value $e_{ex}^l = 2e_{ex}/3$, where we have used the result, $m^*(k_{f_0}, r_0)/m = 0.67$ in SNM and standard value of $E_s(r_0) = 30 MeV$. The results are shown in figure 8 and are compared with the predictions of microscopic DBHF [22,24] and EBHF+3-BF [21] calculations. The momentum dependence of Lane potential extracted from experimental data on nucleon-nucleus scattering upto 100 MeV [50] is also shown in the same figure by the shaded region for purpose of comparison. The results of the present calculations together with those in the microscopic DBHF and EBHF+3-BF calculations have the common feature that all of them pass within the shaded region of Lane potential. It is also noticed that the curves for the choice $e_{ex}^l = e_{ex}/3$ agree reasonably with the microscopic DBHF results of Ref.[22], whereas, the result of the critical case $e_{ex}^l = 2e_{ex}/3$ compares with the EBHF+3-BF calculations within the shaded region.

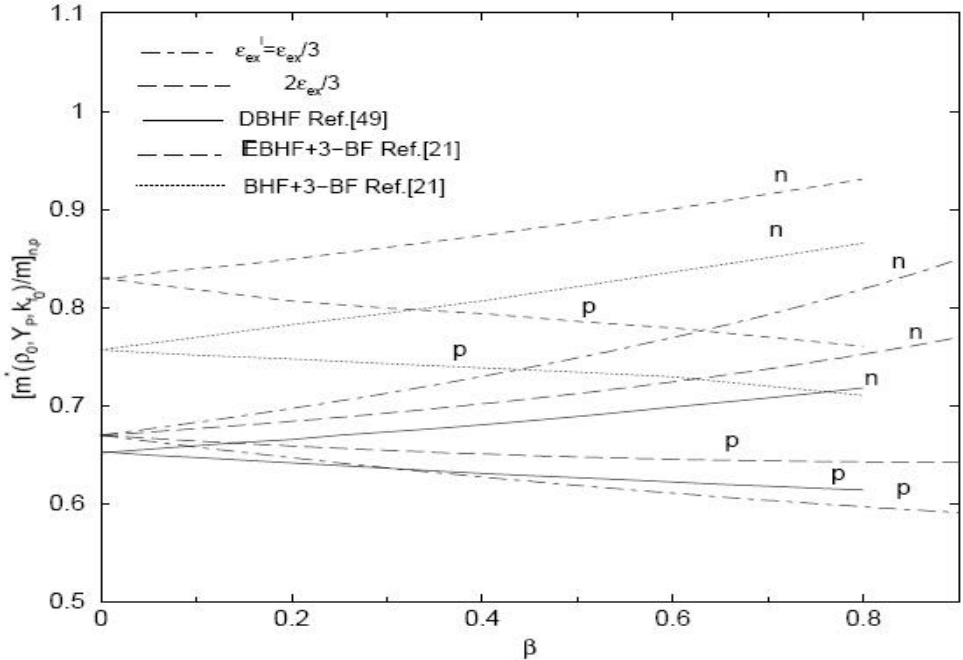

**Figure 7.** Neutron and proton effective masses in neutron rich ANM at zero-temperature, $T = 0$, and normal density, $\mathbf{r} = \mathbf{r}_0$, shown as function of isospin asymmetry $b = (1 - 2Y_P)$ for the two different splittings of exchange strength parameter $\left(e_{ex}^l + e_{ex}^{ul}\right)$, namely, $e_{ex}^l = e_{ex}/3$ and the critical case $2e_{ex}/3$. The results of microscopic DBHF calculations with Bonn B potential [49] (solid line), BHF+3-BF (dotted line) and EBHF+3-BF (dashed line) [22] are also shown for comparison. The ascending curves are for the neutron effective masses where as the descending ones are for proton effective masses.

The thermal evolutions of various properties of SNM and PNM considered in this section require only the knowledge of finite range exchange interactions operating between pairs of like and unlike nucleons (i.e., knowledge of the parameters $e_{ex}^l$, $e_{ex}^{ul}$ and $a$). However we require the knowledge of other parameters, namely, $g$, $b$, $e_0^l$, $e_0^{ul}$, $e_g^l$ and $e_g^{ul}$ for a complete study of mean fields as well as EOSs in both SNM and PNM. The procedure of constraining these additional parameters on the basis of informations coming from saturation properties of SNM, transport model analysis of flow data in high energy heavy-ion reactions, monopole mode of vibrations in finite nuclei and analysis of high density behaviour of asymmetric contribution to the nucleonic part of energy density in NSM have been discussed in detail in Refs.[45,46]. Following this procedure the parameters are determined for the specific case of $g=1/2$ and $e_{ex}^{l}=2e_{ex}/3$, and the values are given in Table 1. The energy per particle at zero-temperature in both SNM and PNM can be obtained as, $e_{0(n)}(r) = H_{0(n)}(r, T=0)/r$. The density dependence of $e_0(r)$ and $e_n(r)$ calculated with the interaction parameters in table 1 are shown in Figure 9. The results of the microscopic DBHF, EBHF + 3-BF and variational calculation using realistic interaction $A18+dv+UIX^*$ [51] are also given in the same figure for comparison. The comparison of the present EOSs are satisfactory with the results of the variational calculation in both SNM and PNM upto a density $r = 0.65\ fm^{-3}$ beyond which the EOSs of the variational calculation show a stiff behaviour. Our results are also in well agreement with the predictions of the EBHF+3-BF calculation for which data upto density $r = 0.5\ fm^{-3}$ in SNM and $r = 0.4\ fm^{-3}$ in PNM are available [52]. Beyond a density $r = 0.25\ fm^{-3}$ the EBHF+3-BF microscopic calculation shows a relatively soft behaviour in SNM. The results of the present calculation agree with the microscopic DBHF results upto density $r = 0.45$ $fm^{-3}$ in SNM and $r = 0.35$ $fm^{-3}$ in PNM beyond which the microscopic DBHF calculation shows a very stiff behaviour, even stiffer than the results of the variational calculation with realistic $A18 + \delta v + UIX^*$ interaction.

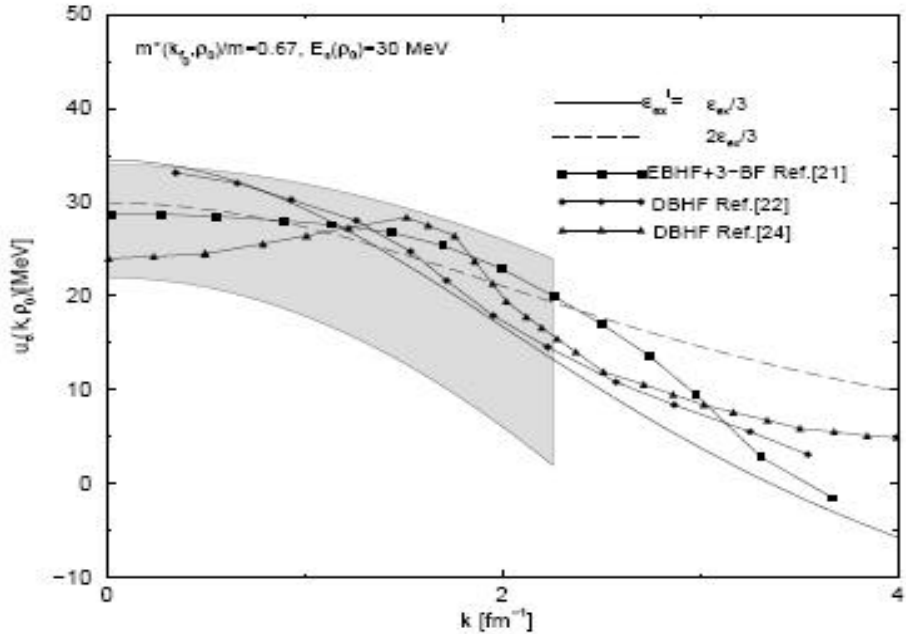

**Figure 8.** $u_t(k, r_0)$ is shown as a function of $k$ for two different splittings of $\left(e_{ex}^l + e_{ex}^{ul}\right)$, namely, $e_{ex}^l$ = $e_{ex}/3$ and the critical case $2e_{ex}/3$. The result obtained in the optical model analysis of the experimental data on nucleon-nucleus scattering upto 100 MeV is shown by the closed shaded area. The results of microscopic DBHF [22,24] and EBHF+3-BF [21] calculations are also shown.

**Table 1:** Values of the interaction parameters appearing in the EOSs of SNM and PNM for the specific case of $g=1/2$ and $e_{ex}^{l}=2e_{ex}/3$.

| $g$ | $e_{ex}^{l}$ [MeV] | $b$ [fm$^3$] | $e_{ex}^{ul}$ [MeV] | $e_{g}^{l}$ [MeV] | $e_{g}^{ul}$ [MeV] | $e_{0}^{l}$ [MeV] | $e_{0}^{ul}$ [MeV] | $a$ [fm] |
|---|---|---|---|---|---|---|---|---|
| 1/2 | -81.230 | 0.5668 | -162.46 | 65.727 | 88.103 | -50.493 | -65.222 | 0.4044 |

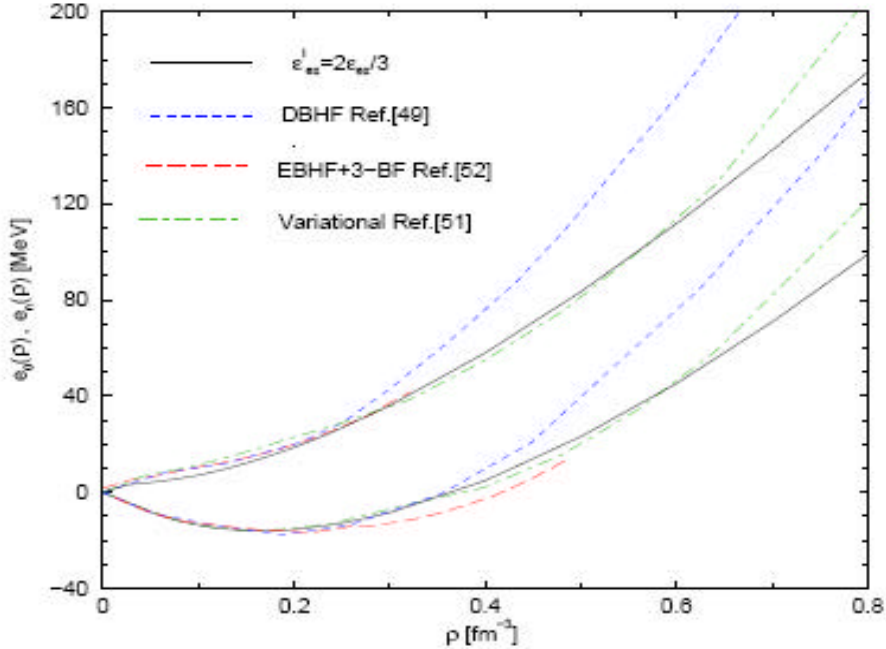

**Figure 9.** Energy per particle, $e_{0(n)}(r_0)$, in SNM (PNM) are shown as functions of $r$ for the specific case of $g=1/2$ and $e_{ex}^{l}=2e_{ex}/3$. The results of microscopic DBHF [49], EBHF+3-BF [52] and variational calculations [51] are also shown for comparison.

## 4. Summary and conclusion

In this paper the EOS of ANM has been formulated in terms of EOSs of SNM and PNM by exploiting the isospin symmetry of various quantities in the EOS of ANM and the fact that SNM and PNM are the two extremes of ANM. The mean field properties as well as the EOSs of SNM and PNM are studied by using a simple finite range effective interaction. The momentum and temperature dependence of the mean fields as well as the temperature dependence of interaction parts of EOSs in SNM and PNM are simulated by the finite range exchange components of the effective interaction operating between pairs of like and unlike nucleons. In the present case the finite range exchange interaction has a Yukawa form having same range $a$ but different strengths $\varepsilon_{ex}^{l}$ and $\varepsilon_{ex}^{ul}$ for interaction between pairs of like and unlike nucleons. It is seen that thermal evolution of various properties of SNM and PNM, i.e., nuclear matter properties relative to their zero-temperature results, can be calculated in this case only in terms of the parameters $\varepsilon_{ex}^{l}$, $\varepsilon_{ex}^{ul}$ and $a$ without requiring the knowledge of other parameters of the effective interaction. The combination $\varepsilon_{ex} = (\varepsilon_{ex}^{l} + \varepsilon_{ex}^{ul})/2$ and range $a$ are determined so as to provide a correct momentum dependence of the mean field in SNM at normal density as demanded by optical model fits to nucleon-nucleus scattering data at intermediate energies, thereby the thermal evolution of various properties in SNM are fixed. In case of PNM there is no such constraint available that could precisely decide the exchange strength $\varepsilon_{ex}^{l}$. However the present status on $(m_n^* - m_p^*)$ splitting in ANM that neutron effective mass goes above the proton one constrains the parameter $\varepsilon_{ex}^{l}$ in between $0$ to $\varepsilon_{ex}$. But the actual value of the parameter $\varepsilon_{ex}^{l}$ is uncertain since different theoretical models widely disagree on the magnitude of $(m_n^* - m_p^*)$ splitting in ANM. In view of this we have studied the thermal evolution of properties in PNM by varying $\varepsilon_{ex}^{l}$ in its range between $0$ and $\varepsilon_{ex}$. Two different behaviours in the thermal evolution of properties, such as, entropy density, internal energy density and free energy density of PNM relative to SNM results have been found which divides the whole range $0$ to $\varepsilon_{ex}$ into two parts. If $\varepsilon_{ex}^{l}$ is in between $0$ and $2\varepsilon_{ex}/3$ thermal evolutions of these properties in PNM will crossover the corresponding curves of SNM at certain high densities. On the other hand, for $\varepsilon_{ex}^{l}$ within $2\varepsilon_{ex}/3$ and $\varepsilon_{ex}$ thermal evolutions of these properties in PNM will remain weaker compared to the SNM results at all values of density and there will be no crossover of PNM curves with that of SNM ones, a behaviour quite similar to those obtained in ideal Fermi gas model. The region to which the actual value of $\varepsilon_{ex}^{l}$ might belong to can be ascertained provided an answer can be obtained to the question whether the entropy density in PNM at a given temperature $T$ can exceed that of SNM. It is important to note that in the ideal Fermi gas model the entropy density in PNM is always below that of SNM at any density. For the critical case $\varepsilon_{ex}^{l} = 2\varepsilon_{ex}/3$, thermal evolution of the properties in PNM approaches the corresponding results of SNM asymptotically at large densities and the magnitude of $(m_n^* - m_p^*)$ splitting in ANM in this case compares reasonably well with the microscopic DBHF results obtained with realistic Bonn potential. In view of the above it may be emphasized that the parameters involved in the finite range exchange components of the effective interactions which govern the momentum dependence of mean fields, like $e_{ex}^{l}$, $e_{ex}^{ul}$ and $a$ in the present case, need to be chosen with care while using the interactions in the studies of temperature dependence of different properties in ANM. The task, in turn, requires more theoretical as well as experimental efforts for a better understanding of the momentum dependence of neutron and proton mean fields in ANM.

**Acknowledgement:** The work is supported by the UGC-DAE Collaborative Research Project of India bearing No. UGC-DAE CSR/KC/2009/NP06/1354 dated 31-7-09. The work is covered under the SAP program of School of Physics, Sambalpur University, India.